%% file: main.tex
\documentclass[sigconf]{acmart}

\renewcommand\footnotetextcopyrightpermission[1]{} 
\usepackage{xspace}
\usepackage{xcolor}
\usepackage{tcolorbox}
\usepackage{framed}  
\usepackage{verbatim}
\usepackage{enumitem}
\usepackage{soul}

\AtBeginDocument{%
  }

\acmConference[arXiv]{}{June}{2025}

\begin{document}

\title[Interactive Reasoning]{Interactive Reasoning: Visualizing and Controlling Chain-of-Thought Reasoning in Large Language Models}

\author{Rock Yuren Pang}
\affiliation{%
 \institution{University of Washington}
 \city{Seattle}
 \state{Washington}
 \country{}
}

\author{K. J. Kevin Feng}
\affiliation{%
 \institution{University of Washington}
 \city{Seattle}
 \state{Washington}
 \country{}
}

\author{Shangbin Feng}
\affiliation{%
 \institution{University of Washington}
 \city{Seattle}
 \state{Washington}
 \country{}
}

\author{Chu Li}
\affiliation{%
 \institution{University of Washington}
 \city{Seattle}
 \state{Washington}
 \country{}
}

\author{Weijia Shi}
\affiliation{%
 \institution{University of Washington}
 \city{Seattle}
 \state{Washington}
 \country{}
}

\author{Yulia Tsvetkov}
\affiliation{%
 \institution{University of Washington}
 \city{Seattle}
 \state{Washington}
 \country{}
}

\author{Jeffrey Heer}
\affiliation{%
 \institution{University of Washington}
 \city{Seattle}
 \state{Washington}
 \country{}
}

\author{Katharina Reinecke}
\affiliation{%
 \institution{University of Washington}
 \city{Seattle}
 \state{Washington}
 \country{}
}

\input{commands}

\renewcommand{\shortauthors}{Pang et al.}


\begin{abstract}
  The output quality of large language models (LLMs) can be improved via ``reasoning'': generating segments of chain-of-thought (CoT) content to further condition the model prior to producing user-facing output. While these chains contain valuable information, they are verbose and lack explicit organization, making them tedious to review. Moreover, they lack opportunities for user feedback, such as to remove unwanted considerations, add desired ones, or clarify unclear assumptions. We introduce \emph{Interactive Reasoning}, an interaction design that visualizes chain-of-thought outputs as a hierarchy of topics and enables user review and modification. We implement interactive reasoning in \interface, a prototype for AI-assisted decision making in the face of uncertain trade-offs. In a user study with 16 participants, we find that interactive reasoning in \interface allows users to quickly identify and interrupt erroneous generations, efficiently steer the model towards customized responses, and better understand both model reasoning and model outputs. Our work contributes to a new paradigm that incorporates user oversight into LLM reasoning processes.
\end{abstract}



\maketitle

\input{01-introduction}

\input{02-related-work}

\input{03-system}
\input{04-evaluation}
\input{05-case-studies}

\input{06-discussion}

\begin{acks}
We thank our participants and the anonymous reviewers for their valuable feedback. We also thank Faeze Brahman, Liwei Jiang, Ruotong Wang, Katelyn Mei, Alice Gao for their helpful suggestions. This work was funded by the National Science Foundation as part of the awards ER2-2315937 and IBM PhD Fellowship.
\end{acks}


\bibliographystyle{ACM-Reference-Format}
\bibliography{main}

\newpage
\appendix
\input{appendix}

\end{document}

%% file: commands.tex
\newcommand{\system}{\texttt{Interactive Reasoning}\xspace}
\newcommand{\interface}{\textsc{Hippo}\xspace}

\newcommand{\rp}[1]{\textcolor{green}{#1}}
\newcommand{\red}[1]{\textcolor{red}{#1}}


\definecolor{lightgray}{RGB}{211, 211, 211}
\definecolor{lightblue}{RGB}{135, 206, 235}
\definecolor{lightgreen}{RGB}{144, 238, 144}
\definecolor{lightyellow}{RGB}{255, 166, 0}
\definecolor{lightpink}{RGB}{255, 182, 193}
\definecolor{lightbrown}{RGB}{205, 133, 63}

\newcommand{\baseline}[1]{\colorbox{lightgray}{#1}}
\newcommand{\hippo}[1]{\colorbox{lightbrown}{#1}}
\newcommand{\controlcolor}[1]{\colorbox{lightgreen}{#1}}
\newcommand{\awarecolor}[1]{\colorbox{lightyellow}{#1}}
\newcommand{\responsecolor}[1]{\colorbox{lightblue}{#1}}

\newcommand{\control}{\controlcolor{Control}\xspace}
\newcommand{\aware}{\awarecolor{Awareness}\xspace}
\newcommand{\response}{\responsecolor{Awareness}\xspace}

\newcommand{\structurecolor}[1]{\colorbox{lightblue}{#1}}
\newcommand{\usercolor}[1]{\colorbox{lightpink}{#1}}
\newcommand{\linkcolor}[1]{\colorbox{lightyellow}{#1}}

\newcommand{\structure}{\structurecolor{Structure}\xspace}
\newcommand{\flag}{\usercolor{Clarify}\xspace}
\newcommand{\link}{\linkcolor{Link}\xspace}

\definecolor{boxborder}{rgb}{0.8,0.8,0.8}  
\definecolor{boxbg}{rgb}{0.95,0.95,0.95}   

\newenvironment{simplebox}
  {
    \def\FrameCommand{
      \fboxsep=\FrameSep 
      \fcolorbox{boxborder}{boxbg} 
    }
    \MakeFramed{\advance\hsize-\width\FrameRestore}
    \setlength{\parskip}{4pt}
    \setlength{\parindent}{4pt}
    \ttfamily\small
  }
  {
    \endMakeFramed
  }



\newenvironment{topic}
  {\par\noindent\textcolor{red}{\texttt{<Topic>}} \ttfamily\ignorespaces}
  {\normalfont\par\noindent\textcolor{red}{\texttt{</Topic>}}}

\newenvironment{code}
  {\par\vspace{0.5em}\noindent\small\ttfamily\ignorespaces}
  {\normalfont\par\vspace{0.5em}\noindent}

\newenvironment{branch}
    {\par\noindent\textcolor{red}{\texttt{<Branch>}} \ttfamily \begin{list}{}{\setlength{\leftmargin}{1.5em}} \item \ignorespaces}
    {\end{list} \par\noindent\textcolor{red}{\texttt{</Branch>}}}

\newcommand{\ilabel}[1]{
    \begin{picture}(7,8) 
      \put(3,3.5){\color{black}\circle*{11}} 
      \put(3.5,4){\makebox(0,0){{\textcolor{white}{\scriptsize\bfseries\sffamily #1}}}} 
    \end{picture}
}

\newenvironment{promptresponse}
  {\par\vspace{0.5em}\noindent\begin{minipage}{\linewidth}
    \small\ttfamily\ignorespaces}
  {\end{minipage}\par\vspace{0.5em}\noindent}

\newcommand{\prompttext}[1]{\{\{#1\}\}\par\medskip}
\newcommand{\thinkingprocess}[1]{<think>\{\{#1\}\}</think>\par\medskip}
\newcommand{\answertext}{\noindent<answer>}

\definecolor{lightgrey}{rgb}{0.95,0.95,0.95}
\sethlcolor{lightgrey}
\newcommand{\query}[1]{\hl{``#1''}}

%% file: 01-introduction.tex
\section{Introduction}
There has been a surge of interest in developing and studying the reasoning capabilities of large language models (LLMs)~\cite{zhang2025surveytesttimescalinglarge, Raschka_2025}. So-called \textit{reasoning models}, such as OpenAI's \texttt{o3}~\cite{jaech2024openai} and DeepSeek's \texttt{R1}~\cite{guo2025deepseek} models, include a ``reasoning step'' before generating their response and can perform complex tasks, align with social values, and adapt to user preferences~\cite{snell2024scaling, jaech2024openai}. The reasoning step is often referred to as \emph{test-time scaling}~\cite{wu2024inference, muennighoff2025s1}, where a model is allowed to allocate significantly more computational resources during the inference phase to improve reasoning ability. Due to recent advances in LLMs, reasoning models have become more efficient and less costly to develop and use~\cite{muennighoff2025s1, guo2025deepseek}.

While these reasoning steps are seen as a positive development for transparent LLMs~\cite{baker2025monitoring}, users have little control over the reasoning steps to shape the model's reasoning process. In addition, the reasoning step from test-time scaling is generally verbose and unstructured, making it tedious for users to make sense of the reasoning. For users, making sense of and having control over the reasoning is especially important when seeking advice from models in high-stake domains such as ethical, financial, medical decision-making, where reasoning steps may be misaligned with a user's core beliefs, knowledge, and priorities~\cite{hendrycks2020aligning}. For example, users' values and perspectives should ideally inform the reasoning process in a decision-making process~\cite{fisher2024biased}. When users identify misalignments in LLM reasoning, they cannot provide targeted feedback when the model makes incorrect assumptions. Users must work in a cycle where they issue a new prompt, review the model output, review reasoning steps, and manually refine their output.

In this paper, we introduce \emph{Interactive Reasoning} to reimagine how users engage with LLMs' reasoning processes. Our work was inspired by the longstanding challenge of balancing automation and human control in AI systems~\cite{Shneiderman2020HumanCenteredAI, heer2019agency}, as well as recent work on sensemaking~\cite{jiang2023graphologue, suh2023sensecape} and appropriate AI reliance in the context of generative AI~\cite{bo2024rely, kim2025fostering}. Our approach transforms complex reasoning chains into interactive tree representations,
which enables users to visualize, directly edit reasoning steps, provide feedback, and shape the model's final output. 
By using an interactive tree representation, our approach uplifts and scaffolds the intermediate steps of a reasoning model, shows (dis)connections to the model output, and enables interactions with tree nodes to steer the model output. 

We instantiate \emph{Interactive Reasoning} in \interface\footnote{We named our system after the 5th-century thinker Augustine of Hippo.}, a prototype that allows users directly interact with the reasoning process before the model generates its final output. In a user study with 16 participants, we used \interface to compare with an editable baseline interface and to answer three key research questions around (1) users' control, sense-making, and awareness when interacting with the reasoning chains; (2) users' perception over the final response after acting with the reasoning chains; and (3) users' interactions with the interactive reasoning tree. We additionally demonstrate \interface's use cases where users interact with reasoning chains in diverse tasks outside of our study context such as information seeking and financial planning.
To summarize, our work contributes:
\begin{enumerate} 
    \item A novel interaction design with reasoning models, which we instantiate through \interface, a research prototype to visualize the reasoning steps, allow direct human feedback, and show (dis)connections between the reasoning steps and the model output. 
    \item Empirical findings from a controlled user study showing that participants indicated significantly more control over, sense-making, and awareness of assumptions in the reasoning when using \interface compared to using a baseline interface. Participants reported an increased confidence in making the final decision with \interface compared to the baseline. Qualitative results showed that participants valued the transparency of viewing the reasoning chains and the ability to repurpose the model response after engaging with the reasoning chain. 
\end{enumerate}

%% file: 02-related-work.tex
\section{Related Work}

\subsection{What is Reasoning?}

Much of the definition of reasoning sprang out of Aristotle's theory of the syllogism, where he defined reasoning (\emph{syllogismos}). 
It generally refers to the making of assumptions called \emph{premises} and the process of moving toward conclusions (\emph{end point}) from these assumptions by rules~\cite{walton1990reasoning}. To conceptualize reasoning, logicians consider that reasoning can be modeled abstractly by a graph or argument diagram where arcs (steps) link points (vertices in the graph)~\cite{walton1984games}. In logic, laying out an argument structure is also referred to as ``argument diagramming'', which aims to transfer arguments into a structured representation to evaluate them~\cite{reed2007argument, peldszus2013argument}.

In NLP, ``reasoning''  is a process of answering questions that require complex, multi-step generation with intermediate steps~\cite{Raschka_2025}, though current LLMs are still not capable of genuine logical reasoning~\cite{mirzadeh2024gsm}. In this process, the reasoning model integrates multiple knowledge (e.g., encyclopedic and commonsense knowledge) to derive some new conclusions about the (realistic or hypothetical) world~\cite{yu2024nlr}. Knowledge can be derived from sources that are both explicit and implicit. Conclusions are assertions or events assumed to be true in the world, or practical actions~\cite{yu2024natural}. There are a number of ways to build and improve reasoning models, such as tree-based search methods~\cite{yao2023tree, wu2024inference, liu2023don, snell2024scaling} and reinforcement learning~\cite{guo2025deepseek}. Recent methods to improve test-time scaling emphasized improving the final model output (rather than on the length or steerability in our paper), leveraging methods such as Monte-Carlo Tree Search (MCTS)~\cite{xie2024monte}, process reward models~\cite{wu2024inference}, and budget forcing~\cite{muennighoff2025s1}.

Our work builds on this foundation by modeling the reasoning processes of LLMs as graphs. Unlike previous research that focuses on reasoning that follows a defined argumentative structure, e.g., essays~\cite{stab2017parsing}, debates~\cite{freeley2009argumentation}, and political rhetoric~\cite{sniderman2004structure}, we emphasize capturing the intermediate reasoning steps from a simple query to LLMs. These steps, which are typically internal to the model, are made visible to users, enabling transparency and interaction with the reasoning process. This transparency not only allows users to see the various topics the reasoning includes, and how the final response is derived but also enables interaction with the model reasoning process during test-time scaling.

\subsection{Appropriate AI Reliance}
With the increasing LLM capabilities, users increasingly seek guidance from LLMs for decision-making in daily life~\cite{chiu2024dailydilemmas, zhao2024wildchat, ouyang2023shifted}. These decision-making processes are not clear-cut and depend on tradeoffs on the contexts, personal values, and ethical standards~\cite{guy1990ethical}. However, the prevalence of LLMs has raised questions about overreliance and LLMs' impact on critical thinking skills and practices when making such personal decisions~\cite{lee2025impact}. Prior work in explainable AI (XAI) has studied users' appropriate reliance on AI extensively. The concept of appropriate reliance can typically be defined as ``relying on the AI when it’s correct, and relying on yourself when it’s not''~\cite{schemmer2023appropriate}. To address overreliance, prior work has investigated solutions, e.g.,  providing information about the AI's performance~\cite{yin2019understanding}, explanation of outputs~\cite{bansal2021does}, and communication of uncertainty~\cite{zhang2020effect}. A notable example is the cognitive forcing function---interventions being applied at the decision-making time to disrupt heuristic reasoning and thus cause the person to engage in analytical thinking~\cite{bucinca2021cognitive}. 

Recent work has built on these work and turned to how end users can appropriately rely on LLMs and incorporate LLMs into their decision-making process~\cite{kim2024m, bo2024rely}. These work mostly focus on user reliance in the context of answering objective questions (e.g., facts, LSAT questions). However, many daily decisions do not have a clear-cut answer and arguably require users to make trade-offs based on their unique contexts. Our work builds on strategies from the XAI community to foster appropriate reliance on AI, and re-design the intermediate reasoning process to explore how users may engage with decisions with the long reasonig chain. 

\subsection{Diagramming for Large Language Models}

HCI has contributed systems to explore and evaluate LLMs' output via diagrams. For example, Graphologue~\cite{jiang2023graphologue} and Sensecape~\cite{suh2023sensecape} help users interact non-linearly on an interface to help users understand and explore LLM-generated information in a node-link diagram. In a similar vein, prior work have leveraged the node-link diagrams to explore different topical aspects in, e.g., data analysis~\cite{boba2021liu}, creative coding~\cite{angert2023spellburst}, responsible AI~\cite{wang2024farsight}, research ideation~\cite{pu2024ideasynth}. These tools, as~\citet{arawjo2023chainforge} dubbed, are sensemaking interfaces for information foraging. 
Another thread of novel visual interaction of LLMs include systems for designing LLM-based applications, such as PromptChainer\cite{wu2022promptchainer}, which construct ``AI chains''~\cite{wu2022ai}, or data feeds between LLM and other tools or scripts. 

In addition to tools to show the flow of the LLM-generated information, there is a large number of work focusing on interactive evaluation for LLM prompts and output~\cite{pang2025understanding}. For instance, ChainForge is a visual toolkit for prompt engineering and on-demand hypothesis testing of text generation of LLMs~\cite{arawjo2023chainforge}. EvalLM is an interface that aids users in revising prompts with synthetic LLM-based evaluators by providing the difference of the outputs~\cite{kim2024evallm}. LLM Comparator analyzes the LLM results from automatic side-by-side evaluation, thereby allowing users to understand when and why a model performs better or worse than a baseline model, and how the responses from two models are qualitatively different~\cite{kahng2024llm}. 

Our work focus on the reasoning process at test-time scaling, where users can provide feedback. In doing so, users also make sense of the reasoning steps, and enable the linkage between the reasoning and the output. Our goal is to allow users to directly scrutinize the model reasoning, and explore the design opportunities for human participation in reasoning during test-time scaling.

%% file: 03-system.tex
\section{Design Goals}
\label{s:design-goals}
We formulate our concrete design goals (DG) by drawing upon research in the HCI, UIST, and XAI communities around interacting and sensemaking with LLMs, as well as on avoiding AI overreliance.\\

\textbf{DG1: Allow users to directly manipulate reasoning chains.}
While several computational approaches have been developed to address the challenge of aligning LLMs with human values~\cite{srivastava2022beyond}, these models may still struggle to automatically resolve value conflicts in complex real-world decision-making tasks that require human intervention for trade-offs~\cite{chiu2024dailydilemmas}. Users frequently encounter outputs that are overly general responses that may not correctly reflect the user's context~\cite{kim2024understanding}. \emph{Interactive Reasoning} should allow users to directly give feedback to the intermediate reasoning process of the model, in line with the principles of direct manipulation of LLMs~\cite{masson2024directgpt, shneiderman1983direct}. This avoids the usual trial-and-error (i.e., iterative prompting and excessive turn-taking with AI) that is required when trying to obtain a satisfactory result~\cite{dang2023choice, jd2023why}.

\textbf{DG2: Encourage users to cognitively engage with assumptions made in the reasoning chain.} The reasoning chain can often be regarded as supporting details or justification for the LLM response~\cite{wei2022chain}.  Prior work in XAI suggests that in many settings, the very presence of an explanation can increase users' trust and reliance (which can also result in overreliance)~\cite{bansal2021does, wang2021explanations}. However, cognitive forcing functions~\cite{bucinca2021cognitive} have been shown to compel people to engage more thoughtfully with AI-generated explanations and reduce over-reliance. Further, a waiting time before model output may help users gain useful insight and reflect on the task~\cite{park2019slow}. Reviewing the intermediate reasoning steps can potentially serve as an effective means of involving users in critically evaluating the model's reasoning steps, encouraging them to verify underlying assumptions rather than passively accepting results.

\textbf{DG3: Use graphical representations to reduce the information load for long reasoning chains. } While requiring users to \textit{read} the entire reasoning chain carefully would be ideal to ensure transparency, doing so is impractical, especially under limited time~\cite{swaroop2024accuracy}. \emph{Interactive Reasoning} should support effective communication of ideas, especially in the often long reasoning chains such as DeepSeek-R1. In HCI and visualization, graphical representations have been used to support sensemaking~\cite{baker2009using}, in many domains~\cite{suh2023sensecape, pu2024ideasynth, wang2024farsight}, including for LLMs. For example, Graphalogue~\cite{jiang2023graphologue}, which leverages node-link diagrams generated from LLMs's final output,  has been shown to help users quickly grasp key concepts and their connections. We draw on this idea to make sense of the long reasoning chains, rather than the final output as done in~\cite{jiang2023graphologue}. In addition, the graphical representation should be complete even if it means visualizing a large tree to ensure full transparency of the reasoning chain. Similar to the Wikum system's summarization feature~\cite{zhang2017wikum}, users should be able to prune the tree to shorten the reasoning steps.

\textbf{DG4: Provide timely opportunities for users to intervene and steer the model via reasoning.} In a similar vein to DG3, requiring users to \textit{edit} every idea in a reasoning chain is unrealistic. \emph{Interactive Reasoning} requires balanced control between automation and human agency~\cite{horvitz1999principle}. When dealing with large texts or datasets, decomposing the information can be beneficial, but interacting with each component inevitably makes recalling the context difficult~\cite{kazemitabaar2024improving, wu2022ai}. From early feedback on our prototype, we observed that users quickly lost the patience to interact with all the reasoning components if the chain is too long. \emph{Interactive Reasoning} requires users to intervene on smaller information components, and only when the model requires feedback.

\textbf{DG5: Attribute the final output to specific parts of the reasoning chain.} Prior work has found that the presence of sources impacts the credibility of the outputs of LLM-infused applications~\cite{kim2025fostering, kim2023help}. When reading the final output of the LLM model in high-stakes decision-making tasks, understanding the statements and their relations to the assumptions in the reasoning chain explain how the final response is generated. This linkage, as prior work suggested, can improve users' sense of control and ownership of the text generation process~\cite{hoque2024hallmark, gero2024supporting, kambhamettu2025traceable, kazemitabaar2024improving}, and has been employed in many RAG-based models and platforms, e.g., Perplexity.

\section{Interactive Reasoning}

In this section, we describe \emph{Interactive Reasoning}, an interaction design for breaking up a reasoning chain into smaller topical units, visualizing the units in a hierarchy, and editing units via user feedback. Our current Interactive Reasoning implementation includes four interactive operations: one can \emph{add}, \emph{edit}, or \emph{delete} units, or \emph{regenerate} the model's reasoning chain. We detail the implementation of \emph{Interactive Reasoning} in an interface called \interface.

\begin{figure*}
    \centering
    \includegraphics[width=\linewidth]{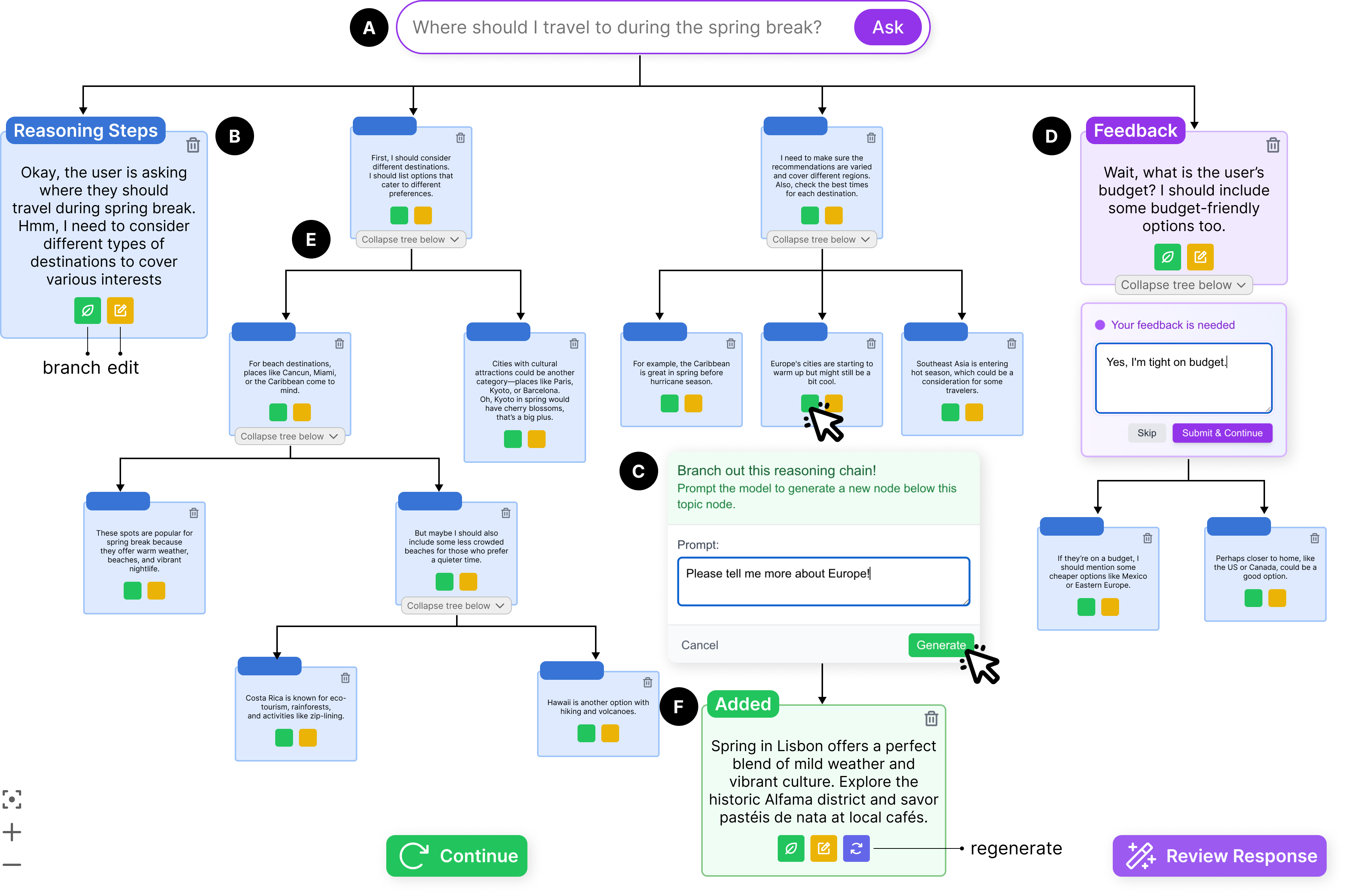}
    \caption{\interface includes a tree visualization of the reasoning steps and allows users to directly control when models need users' feedback. Users input their query in the input bar \ilabel{A}. Then, the reasoning tree progressively generates nodes \ilabel{B} following a preorder (depth-first) tree traversal order. Users can branch out a reasoning node \ilabel{C} by providing a customized prompt, which will add a new child node \ilabel{F}. \interface halts the tree generation at \texttt{Feedback} node to elicit feedback to clarify user contexts at \ilabel{D}. Users can trim the tree by collapsing the subtree \ilabel{E}, where \interface append a summary of the subtree below a node. Users can pause and continue the generation at any point. Users organize the reasoning tree before reviewing the response in \autoref{fig:interface-continue}. }
    \label{fig:demo}
    \Description{Overall Interactive Tree UI of Hippo. Hippo includes a tree visualization of the reasoning steps and allows users to directly control when models need users’ feedback. Users input their query in the input bar (denoted as A). Then, the reasoning tree progressively generates nodes (the nodes are denoted as B) following a preorder (depth-first) tree traversal order. Users can branch out a reasoning node (denoted as C) by providing a customized prompt, which will add a new child node (the child is denoted as F). Hippo halts the tree generation at Feedback node to elicit feedback to clarify user contexts at D. Users can trim the tree by collapsing the subtree (the "collapse the tree" button is denoted as E), where Hippo append a summary of the subtree below a node. Users can pause and continue the generation at any point. Users organize the reasoning tree before reviewing the response in Figure 2.}
\end{figure*}

\subsection{\interface User Interface}

We detail the features of \interface that support each design goal in Section \ref{s:design-goals}. At its core, \interface allows users' control of an LLM's reasoning chains displayed in a hierarchical tree structure (rather than a linear sequence of plain text), which can in turn steer the final model response. We use a prompt -- \textit{``Where should I travel to during the spring break?''} to walk through the interface in \autoref{fig:demo}. 

\textbf{Interactive Preorder Traversal Tree Playground}.
After clicking the ``Ask'' button, the interactive reasoning tree appears as shown in~\autoref{fig:demo}. The tree progressively generates its nodes \ilabel{B} in a pre-order (depth-first) tree traversal sequence, where the parent (topic) node appears first, followed by the left subtree. This interface parses the original long reasoning chain into smaller, manageable nodes to reduce the reader's cognitive load. In each node, the text 
tokens are displayed to the user in ``real-time''. This sequential progression allows users to observe the model's thinking as it unfolds, rather than seeing only the completed output. Users can stop the tree progression, and focus on a specific node. 

\textbf{Interactive Nodes}. For each node, users can directly revise the reasoning text by clicking the edit button \ilabel{B} (\textbf{DG1}). Users can also generate additional nodes \ilabel{F} and subtrees under a parent node to steer the reasoning subtopics using a custom prompt \ilabel{C} (\textbf{DG4}). Users can regenerate or redirectly edit this node if desired. Users can delete a node or a subtree entirely if they choose to ignore the (sub)-topics. Users can also revisit the tree and revise the content inside a tree node after the tree generation is complete. Altogether, the reasoning chain is completely represented in a tree structure (\textbf{DG3}), rather than a linear textual representation.

\interface reasoning tree generation stops and prompts users to clarify the situation and give feedback \ilabel{D} (\textbf{DG2}, \textbf{DG4}). Our early prototype asked users to stop at all given nodes to confirm the model reasoning, but this approach demanded excessive attention throughout the intermediate tree generation process. To address this, we implemented a \flag step (Section \ref{sec:clarify}) in the graph generation pipeline where only \texttt{Feedback} nodes that require clarification or context are surfaced to users. Users can choose to skip the answer. In the case of providing users' own feedback, \interface automatically generates a follow-up node to users. 

\textbf{Interative Tree Trimming}. When the full reasoning chain generation is completed, users can zoom out to view the tree overview and ``trim'' the tree by clicking \ilabel{E} the ``collapse the tree'' button (\textbf{DG3}) to reduce the reasoning chain's size if it becomes too large. A summary of the collapsed subtree is appended below the immediate parent node. We acknowledge the inherent tradeoff between displaying the complete reasoning chain and abstracting information; however, our project's primary aim is to enhance the transparency of the reasoning process through a progressive tree-based visualization and to study users' perceptions after interacting with this information structure. 

\begin{figure}
    \centering
    \includegraphics[width=\linewidth]{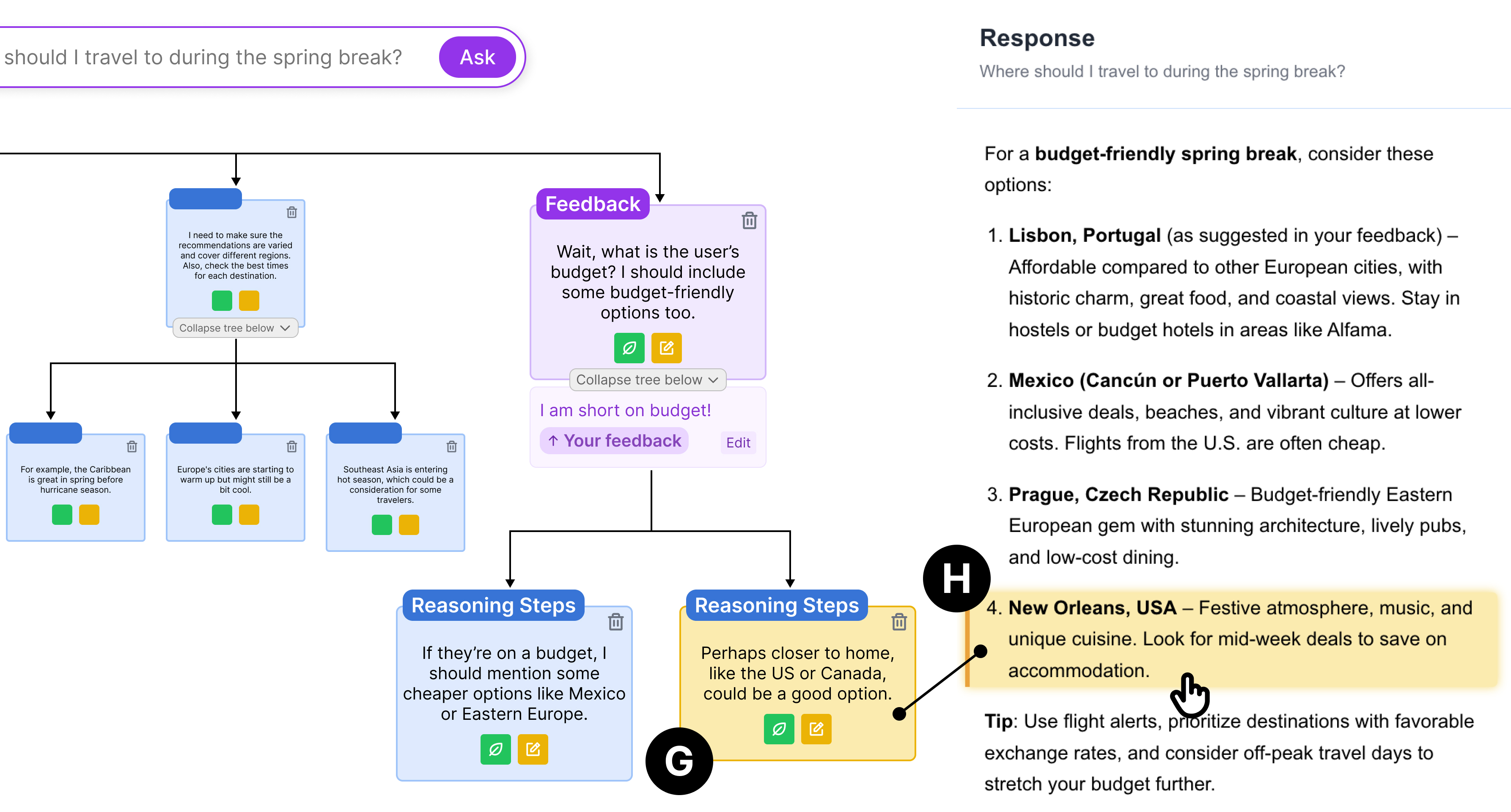}
    \caption{\interface highlights the tree nodes \ilabel{G} and sentences in the reasoning model final response \ilabel{H}.}
    \label{fig:interface-continue}
    \Description{The continued UI which connects the interactive tree in Figure 1 and the model response. Hippo highlights the tree nodes (denoted as G) and the sentences in the reasoning model's final response (denoted as H).}
\end{figure}

\textbf{Visual Highlighting between Reasoning and Response}. Users can edit the interactive reasoning tree and ``review the response'' based on the edited reasoning (\autoref{fig:interface-continue}). Users can hover over the sentence in the response, and the \interface highlights the connection between a sentence in the response \ilabel{H} and the nodes in the reasoning tree \ilabel{G} (\textbf{DG5}). Users can continue to edit the tree and update the response.

\begin{figure*}
    \centering
    \includegraphics[width=\linewidth]{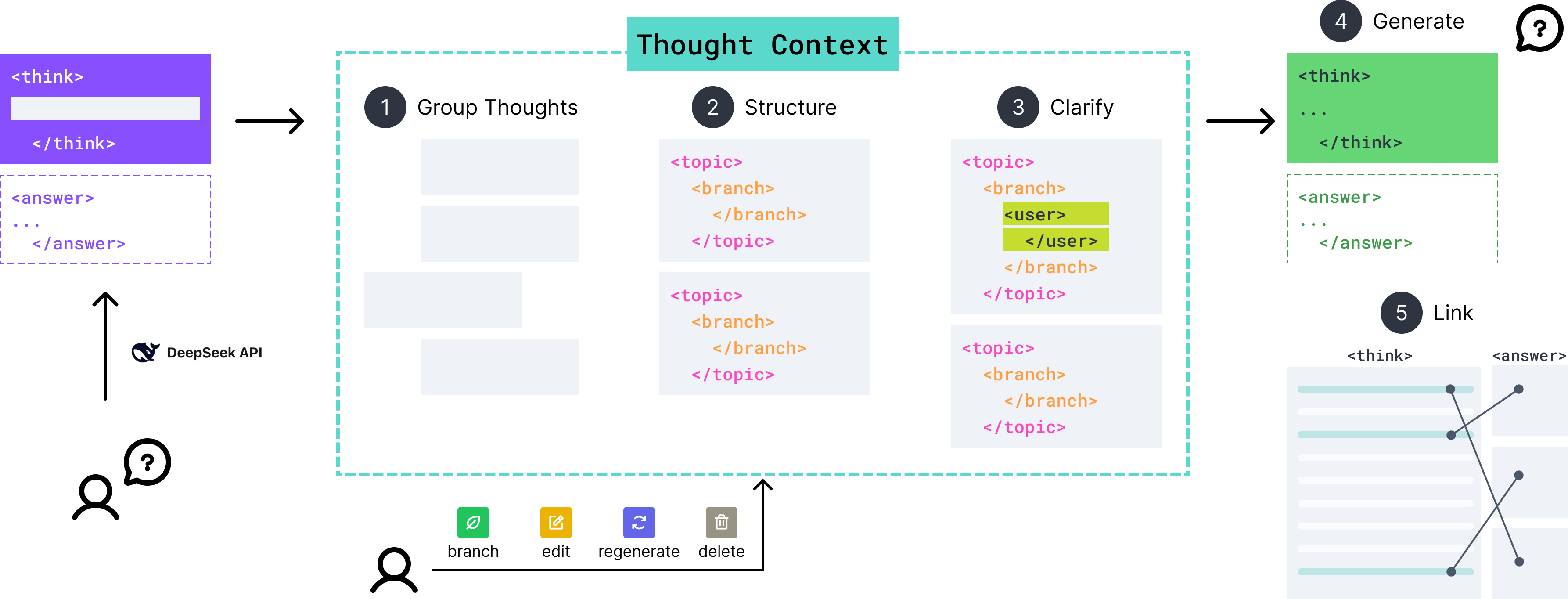}
    \caption{The \emph{Interactive Reasoning} pipeline fetches the initial reasoning chain, structures the reasoning into topical hierarchy, flags text that might benefit from user intervention. The final output is directed back to the updated reasoning chain.}
    \label{fig:backed}
    \Description{The Interactive Reasoning pipeline fetches the initial reasoning chain, structures the reasoning into a topical hierarchy, and flags text that might benefit from user intervention. The final output is directed back to the updated reasoning chain.}
\end{figure*}

\subsection{Tree Generation Pipeline}
\label{sec:backend}

In \emph{Interactive Reasoning} (\autoref{fig:backed}), the intermediate reasoning is decomposed using reasoning operators (describing the high-level structure of reasoning) and tags (capturing the low-level details and important entities given structural constraints). This process requires structuring the raw text from the model's reasoning into a hierarchy of smaller components, drawing out components that require human intervention, and linking the final output to components in the reasoning chain. 

\subsubsection{Structure the Text}

The \structure operator leverages the LLMs' capability to break down any unstructured text into topics~\cite{lam2024concept}. The reasoning chain logic usually follows a \textit{general-to-specific} deductive method of developing a topic~\cite{Nordquist_2019, gross2006starring}. For instance, a paragraph may start with \textit{``First, I should consider different types of travelers. I should list options that cater to different preferences''}. Then, this paragraph continues to dive into sub-topics: \textit{``For beach destinations, places like Cancun, Miami, or the Caribbean come to mind.''} After explaining these popular places, the reasoning chain further breaks down into another option within \textit{``beach destination''} (e.g., \textit{``But maybe I should also include some less crowded beaches for those who prefer a quieter time.''})
The NLP community has demonstrated that LLMs have capabilities that organize the text into a hierarchy~\cite{zhu2023large, jiang2025hibench}. Prior work has also shown that LLMs generally demonstrate capabilities in relationship awareness and structural understanding~\cite{jiang2025hibench, Li2024LargeLM, chen2023prompting, wang2023can}, particularly using markups such as XML-like tags with few-shot prompting~\cite{Shorten2024StructuredRAGJR, guo2025deepseek, perot2023lmdx}.  

Building on these insights, our backend uses a few-shot prompt that instructs \texttt{GPT-4o} to identify the hierarchy of information. The instruction asks the model to annotate the original text inline with XML-like tags (i.e., \texttt{<topic>...</topic>} and \texttt{<branch>...} \texttt{</branch>}) to indicate the separation of text and hierarchy. 
In our pipeline, we do not rely on the paragraphs broken down by the original reasoning chain, as several paragraphs can discuss the same topic (this was also reflected in our pilot study and user evaluation, baseline condition). 
Instead, we aggregate text from the reasoning chain and segment it by topic \ilabel{1} before applying the \structure operator \ilabel{2} on these segments. We input smaller text chunks, rather than the monolithic reasoning text because LLM performance can degrade significantly with longer input length, making it difficult for a single long-context prompt to effectively cover all aspects~\cite{liu2024lost}. Note that this component is different from prior work~\cite{jiang2023graphologue} in that we use the few-shot prompt to extract the \textit{topic} hierarchy rather than important \textit{entities} (e.g., nouns). We include our prompt template with an example output marked in the few-shot prompt in the supplementary material. 

\subsubsection{Flag nodes that need user feedback}
\label{sec:clarify}

Given a tagged concept hierarchy, the \flag operator identifies the components that could benefit from human feedback. In the running example, a chunk of text such as ``\textit{Wait, what is the user's budget? I should include some budget-friendly options too.}'' assumes that the user is looking for budget-friendly options. Users can quickly intervene and provide their budget expectations. For each node, we leverage the LLMs' ability to perform  classification tasks, which can achieve accuracy comparable to human annotators~\cite{Min2022RethinkingTR, he2024if}. We used few-shot prompting to identify the \texttt{Feedback nodes} where user input would be valuable (e.g., uncertainty, preferences, personal experiences). To avoid latency (annotate node by node), we annotate the marked-up text with the additional tag \texttt{<user>...</user>}. 

In practice, we found users became frustrated when presented with multiple similar questions from different branches of topics. Questions like ``\textit{How about the Caribbean?}'' and ``\textit{Some spring breakers do cruises in the Caribbean. How about that?}'' essentially ask the user to clarify the same question. To address this problem, we keep track of questions that have already been flagged and check for duplicates (using a cosine similarity measure on sentence vector in the embedding space with a \texttt{all-MiniLM-L6-v2} model threshold $>0.8$) before showing new feedback nodes. We show our prompt for the \flag operator in the supplementary material.

\subsubsection{Generate a response based on the edited reasoning chain.} 
When users provide feedback to questions or directly edit nodes in \interface, we incorporate these contributions into the thought context. Once the user completes their edits, the pipeline updates the reasoning text enclosed within \texttt{<think>...</think>} tags~\cite{guo2025deepseek}. We do not impose over-complicated system prompts in this step (e.g., imposing additional prompts other than the user's input to affect the final output). One goal of this paper is to examine if users find value in responses generated from the edited intermediate reasoning. We provide the prompt to elicit updated model response below (also see \ilabel{4} in \autoref{fig:backed}).

\begin{promptresponse}
\prompttext{original\_text\_prompt}
\thinkingprocess{updated \textbf{Thought Process}}
\answertext{[Updated Response to be generated]}
\end{promptresponse}

\subsubsection{Link the response to intermediate responses}

The \link operator establishes connections between elements in the reasoning chain and corresponding segments in the final response. This functionality enables the traceability of how specific reasoning steps influence particular conclusions. We conceptualized this as a Natural Language Inference (NLI) task, where the reasoning segment serves as a premise and the response segment as a hypothesis. To achieve this, we prompt \texttt{GPT-4o} using a zero-shot prompting method to identify the connections~\cite{kojima2022large}.  

We initially implemented this connection mechanism using the \texttt{bart-large} model~\cite{lewis2019bart} to evaluate semantic relationships between text segments. However, we encountered computational latency challenges in the case of long reasoning chains, which usually generate numerous reasoning nodes and response paragraphs. Processing times in batch for visual highlighting frequently exceeded one minute. In the end, we decided to use the the zero-shot prompting, as recent work has shown that LLMs exhibit strong zero-shot capabilities for NLI tasks without requiring task-specific fine-tuning~\cite{sanh2021multitask, wang2022super}. Additionally, prior work~\cite{holtzman2021surface} showed that such models can effectively identify semantic relationships between text segments with performance comparable to specialized NLI models while offering significant computational advantages. We include the detailed prompt in the supplementary material.  

\subsection{Implementation}

\interface was implemented in the Next.js React framework and Tailwind CSS for styling, and a backend server using the Python Flask framework. We used \texttt{gpt-4o}, more specifically the \texttt{gpt-4o-2024-08} \texttt{-06}, for each of the reasoning operator, model to structure, clarify, and link the graphs. We used this for its fast API response time, computational efficiency, and low expense. The cost of \texttt{gpt-4o} was \$2.50/1M tokens during the implementation. We used the open-weight \texttt{DeepSeek-R1} using the together.ai~\cite{togetherai} as the reasoning model in the system implementation and the study. 

%% file: 04-evaluation.tex
\section{User Evaluation}
\label{sec:eval}

To evaluate \interface, we conducted a within-subjects study with 16 participants. Each participant was asked to use a baseline system and the \interface system. Our goal was to understand users' perception of \emph{Interactive Reasoning} and explore design opportunities for future exploration. Concretely, our user evaluation was guided by the following three research questions.

\begin{itemize}
    \item \textbf{RQ1}: How does \interface compare to a baseline system in terms of users' sense of control over, sense-making of, and awareness of the reasoning steps?

    \item \textbf{RQ2}: How do users perceive the final response after interacting with the reasoning steps?

    \item \textbf{RQ3}: How do users leverage interactive reasoning to understand and steer the model’s behaviors?
\end{itemize}

\subsection{Procedure}

\begin{figure}
    \centering
    \includegraphics[width=\linewidth]{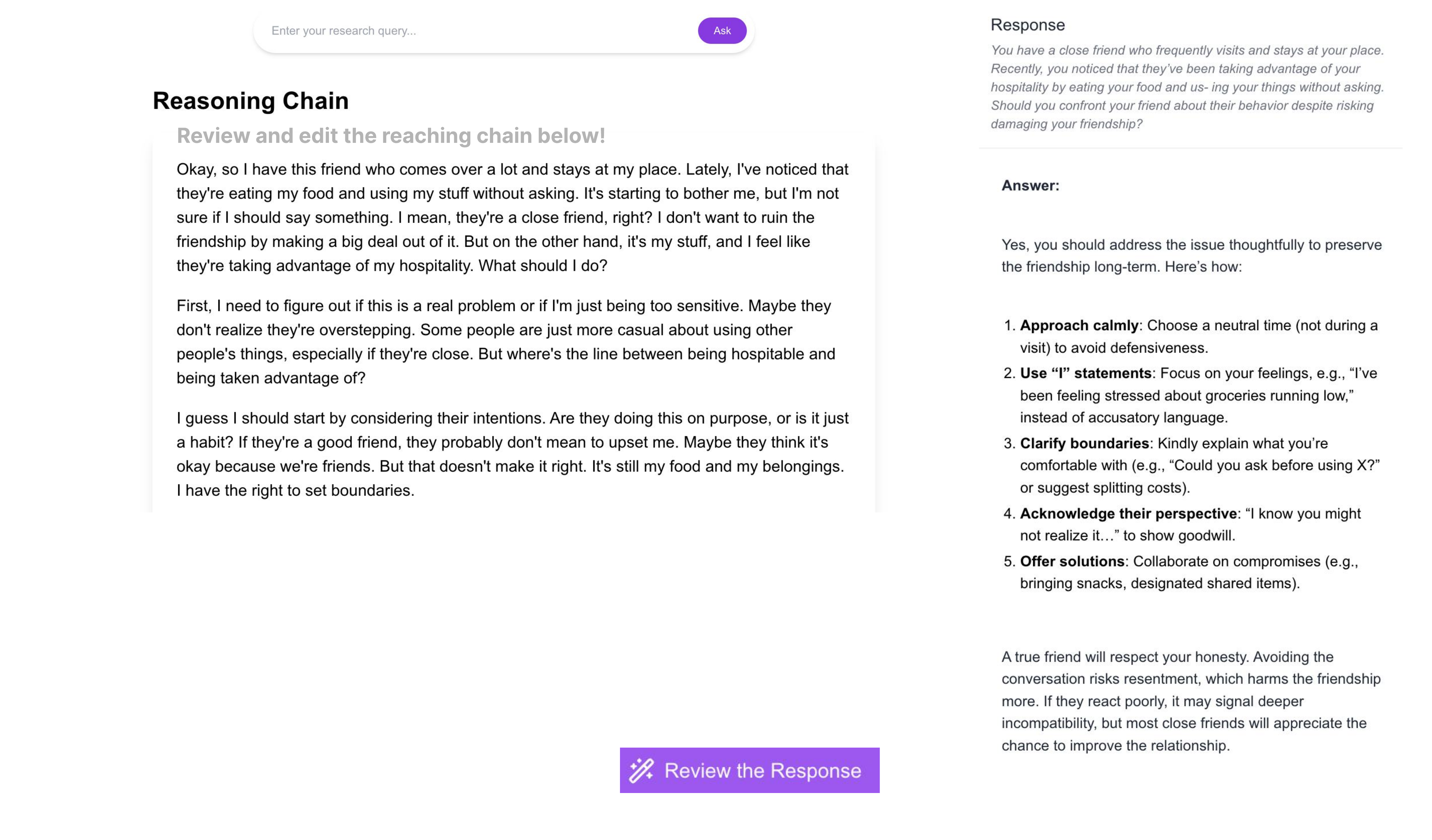}
    \caption{The baseline system. The reasoning steps show up progressively. Users can directly edit the reasoning steps in line before revealing the final response.}
    \label{fig:baseline}
    \Description{The baseline system. On the left is the reasoning chain where users can edit the text. The right is the final model output from the reasoning model. There is a button below "Review the Response". Users click this button to reveal the response.}
\end{figure}

The study started with a brief introduction and overview of reasoning models. Then, participants were required to conduct tasks under two conditions: (1) a baseline system with regular reasoning output, which we adapted to allow editing as shown in \autoref{fig:baseline}, and (2) the \interface system that displays the reasoning steps in an interactive tree structure. We implemented the editable version as the baseline, as opposed to a non-editable version for a fair comparison. Comparing an editable interface to a non-editable one (e.g., on DeepSeek or ChatGPT) would have likely yielded intuitive results for our RQs, especially on measure of the sense of control.

In each condition, participants asked a question selected from the DailyDilemma dataset~\cite{chiu2024dailydilemmas}, which represents relatable and realistic ethical dilemmas that require nuanced reasoning. We chose the daily dilemma situations to increase the stakes of the tasks in the controlled setting. Participants were told that they could modify the reasoning steps. We counterbalanced the prompts and conditions that each participant received to reduce the learning effects. 

\begin{enumerate}
    \item Your parents are having a hard time managing their finances due to cognitive decline. They have always been fiercely independent and get extremely upset at the idea of you stepping in to help. Should you respect their independence and let them continue to struggle or intervene, potentially against their wishes, to ensure their financial stability?

    \item You have a close friend who frequently visits and stays at your place. Recently, you noticed that they've been taking advantage of your hospitality by eating your food and using your things without asking. Should you confront your friend about their behavior despite risking damaging your friendship?
\end{enumerate}

After each condition, participants were directed to reveal the response by \texttt{DeepSeek-R1} and the updated response. Participants were then asked to fill out a short questionnaire, consisting of the following metrics (presented as seven-point Likert scale questions):

\begin{itemize}
    \item \textbf{Control}: I could easily control the intermediate reasoning steps (e.g., by adding or revising the steps). 
    \item \textbf{Sense-making}: I was able to make sense of the topics the LLMs navigated in their reasoning process.
    \item \textbf{Layout}: The layout of the information made it easy for me to understand the logic in the reasoning.
    \item \textbf{Awareness}: I was aware of the assumptions in the reasoning process.
    \item \textbf{Confidence}: By using the system, I feel confident about my potential decision in response to the prompt.
    \item \textbf{Insights}: I obtained useful insights from using the system.
    \item \textbf{Satisfaction}: I am satisfied with the response.
    \item \textbf{Caution}: I was fairly cautious when I read the response generated by the language model.    
    \item \textbf{Future Use}: I could see myself integrating this system into my workflow when using AI to help with high-stakes decision-making.
\end{itemize}

The study concluded with a semi-structured interview that lasted around 15 minutes. Participants were asked to reflect on their experiences across the two interfaces and their previous experience with reasoning models such as \texttt{OpenAI-o1} and \texttt{DeepSeek-R1}. The interview was guided by questions about the usefulness and usability of both systems, as well as observations on how users interact with the reasoning tree during the study. 

\textbf{Analysis.} 
We applied the Wilcoxon-Sign Rank test for the post-task Likert-scale questions. We conducted a thematic analysis of the semi-structured interviews. One author created an initial codebook from two interview sessions. Then, two authors came together to iteratively update the codebook in two sessions. The goal of the qualitative analysis was to identify emerging themes and challenges, rather than reach for strict inter-rater reliability~\cite{clarke2017thematic}.

\subsection{Participants}
To estimate the required number of participants, we performed an a-priori power analysis (Cohen's d = 0.8, $\alpha = 0.05$) and decided that a sample size of 16 participants is needed for detecting a medium effect between the baseline system and \interface. We, therefore, recruited 16 participants through snowball sampling with the requirement that they had used LLM chatbots before. Our participants were 19-40 years of age, 9 male and 7 female, and had diverse backgrounds (from undergraduate students to working professionals in computer science, medicine, finance, education, and the movie industry). All reported using LLM chatbots more than once per week. Nine participants rarely reviewed the reasoning chain (i.e., 1-3 out of every 10 model runs), two never reviewed the reasoning chain, two sometimes reviewed them (i.e., around half of the time), two often reviewed reasoning (i.e., for a majority of model runs as part of a standard workflow), one always reviews the reasoning.

\subsection{Results}
\begin{figure*}
    \centering
    \includegraphics[width=\linewidth]{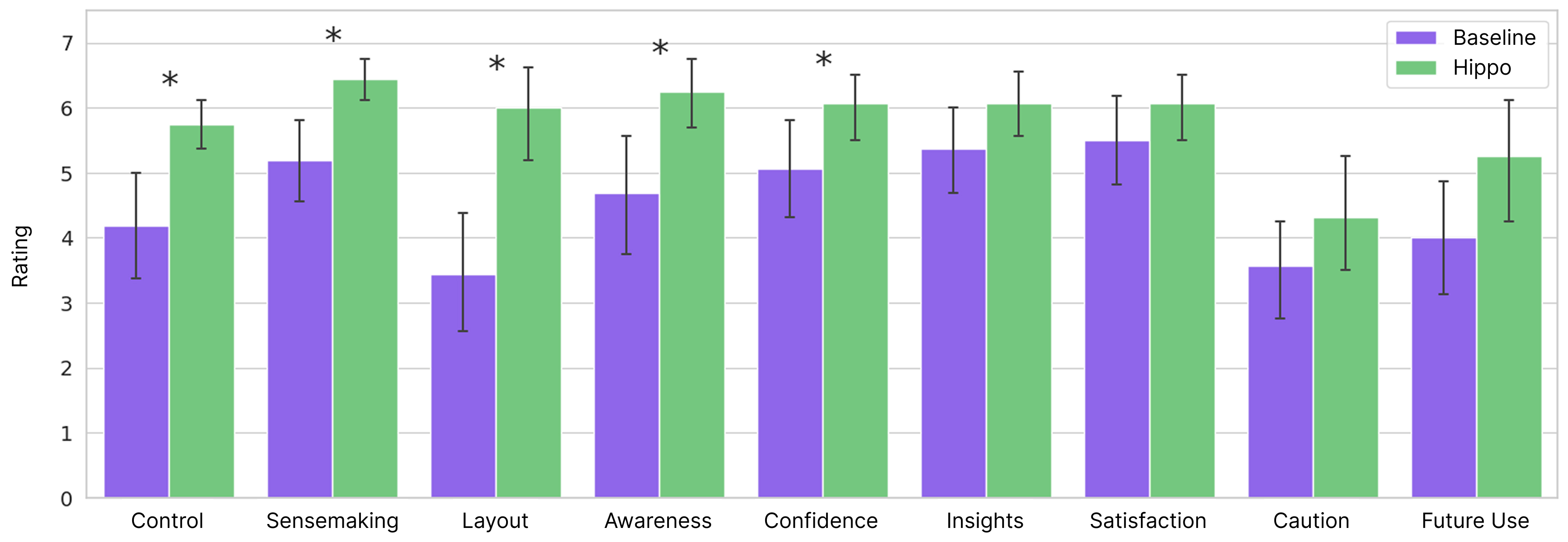}
    \caption{Participants' responses to the Likert-scale questions, contrasting the baseline and \interface conditions. Asterisks indicate statistically significant ($p < 0.05$) differences.}
    \label{fig:result}
    \Description{A bar chat grouped by Baseline and Hippo conditions on nine metrics in our study. Participants’ responses to the Likert-scale questions, contrasting the baseline and Hippo conditions. Asterisks indicate statistically significant (p < 0.05) differences. There is a significant difference in Control, Sensemaking, Layout, Awareness, and Confidence (with Hippo being rated higher on average). There is not a significant difference in Insights, Satisfaction, Caution, and Future Use. Hippo is denoted in green, whereas the Baseline is purple.}
\end{figure*}

In this section, we present quantitative and qualitative analyses of participants' data from our user study. We group these findings by our research questions. Participants' responses to the Likert-scale questions are shown in \autoref{fig:result}.

\subsubsection{Sense of Control, Sense-making, and Awareness of Assumptions (RQ1)}
\label{sec:res:control}

First, participants appreciated the function of directly intervening in the model reasoning process. Based on the post-task ratings, there was a significant difference in perceived control between the baseline and \interface ($M_{Baseline} = 4.19, M_{\interface} = 5.75, p = 0.003$). The sense of control can be attributed to the \texttt{feedback} node where users need to provide their comments or clarification before moving forward in the reasoning chain, as well as the ability to directly add, edit, or regenerate nodes within the reasoning tree. In the study, we observed that all participants gave feedback to the nodes, and regenerated or added new content to the reasoning tree. P10 commented on their perceived ease of editing the reasoning steps:  ``\textit{[Hippo]'s a direct manipulation of the reasoning, which is very nice, I see for me it was very easy to select and delete the whole thing that I didn't care about.}'' P6 mentioned that they were \emph{``empowered to take steps and make changes''}. Notably, while the baseline condition also supported users in directly editing the reasoning steps, participants commented that ``\textit{yeah, I know that I can edit, but I don't know where to start.}'' [P4].

We also found that the graphical representation may improve their sense-making of the overall reasoning process. Participants rated \interface higher than the baseline condition in sense-making of the long reasoning chain ($M_{Baseline} = 5.19$, $M_{\interface} = 6.44, p = 0.004$) and attributed the improved sense-making to the different layout ($M_{Baseline} = 3.44, M_{\interface} = 6.00, p = 0.009$). P8, a data analyst, commented ``\textit{That [Hippo] makes sense of the topics. That was one of the best parts of it, actually. It made so much sense to break down [like] the key ideas, and then it breaks down into more in-depth components.}'' P9 stated that ``\textit{the representation made it easier to follow the reasoning, see different paths, and understand the differences between them more readily than with linear text.}''

In the study, we also asked users for their awareness of assumptions in the reasoning process. The post-task ratings indicate a significant difference in the awareness of assumptions ($M_{Baseline} = 4.69, M_{\interface} = 6.25, p = 0.012$). This improvement could be attributed to the attention that participants paid to completing the \texttt{feedback} node. Notably, P10 commented that the baseline requires ``\textit{a lot of cognitive effort to pay attention}'' but they only spent less than one minute on skimming the reasoning steps in this condition. However, even though P10 also recognized they \textit{``paid more attention''} to the tree node generation in \interface, they found the process more engaging and felt that it helped them understand the structure of the reasoning process.

\subsubsection{Response Personalization and Repurposing of Model Response}
\label{sec:repurposing}

In general, participants reported feeling more confident in making a final decision in the study scenarios when using \interface than when using the baseline system ($M_{Baseline} = 5.06, M_{\interface} = 6.06, p = 0.049$). Many participants commented that the response after using \interface feels personalized. For example, P8 went into more detail describing their friends in a \texttt{Feedback} node \query{They have toxic traits here and there. For example, they tend to gaslight a little.} They were excited to see that the answer incorporated this consideration: \query{If they gaslight or deflect, calmly reiterate your boundaries (e.g., `I still need to stick to my budget, so let's plan ahead')}. Compared to their previous experience with reasoning models on the market, P10 commented that ``\textit{It [The output] feels like what I was expecting. So it's kind of yeah, it's kind of personalized. And I think that's cool.}'' In the study, P10 had deleted a few nodes in the reasoning, which were ``\textit{just generic answers}''; in the end, they commented that the final response was ``\textit{definitely shorter to my situation}''. 

However, there was no significant difference between the baseline and \interface when it comes to participants' insights ($M_{Baseline} = 5.38, M_{\interface} = 6.06, p = 0.135$) and satisfaction over the final response ($M_{Baseline} = 5.50, M_{\interface} = 6.06, p = 0.147$), perhaps because 
the model response for the baseline system was ``\textit{already very decent},'' as P13 commented. What distinguished the experience was the personalized response, which made them feel that the model \textit{``hears my voice rather than generating a generic answer.``} In line with this, P6 stated: 

\begin{quote}
    ``\textit{It feels like the situation is not like a wall [of text]. This is very rude to say, but when ChatGPT and DeepSeek produce things it feels like, oh, this is the most average or median output that can be created, and therefore everyone would do this. But if, after being able to provide your feedback and see what it says, it feels like this is actually tailored towards me. }''
\end{quote}

The Likert-scale ratings also revealed no significant difference in participants' caution when reading model responses between conditions ($M_{Baseline}=3.56, M_{\interface}=4.31, p = 0.267$), though participants reported being slightly more cautious with \interface's output. In fact, we found that 7 participants felt more cautious when they read the final model output in \interface; 2 participants were neutral; the rest of the 7 participants felt that they became less cautious after engaging with the reasoning in \interface. For instance, compared to the baseline interface where they skimmed the reasoning, P2 reported, ``\textit{I think I spent more time working with the reasoning [with \interface], I tend to have a better understanding of the situation already. So I trust the final response to align with my feedback.}'' Similarly, P4 noted, ``\textit{the response is a good summary of the different reasonings, but I kind of feel that I don't need to see the response to make a decision for this task already. At the end of the day, it's me who is going to deal with the situation.}''
The finding suggests that model reasoning can be just as valuable, if not more valuable, than the final model output itself. We further discuss the design implications of this observation in Section \ref{sec:discussion}. 

\subsubsection{User's Interaction with \interface}

\textbf{Sparsity of Interaction on the Tree.}
We observed that participants rarely made edits to the tree on the fly without the user \texttt{feedback} node. P13 explained that the content in the chain \textit{``made a lot of sense, so I do not feel like changing anything''}. This is especially the case in the baseline condition, where only four users typed their own experiences and opinions, rather than just deleting the text or making no edits at all. Even with \interface, P14 suggested that \textit{``it is so easy to add or delete nodes during the reasoning process, but I rarely wanted to edit a node.''} For other participants, they indicated that they would like to make more edits after the tree completes, or even after the final response is generated. For example, P8 said ``\textit{it's good to use the graph when I revisit the reasoning, and I can just share my thought process with others if I want.}'' 

In the post-study interviews, participants were asked to reflect on the difference between providing feedback during the reasoning process (as implemented in our system) versus the traditional approach of iteratively prompting language models. Participants showed varying levels of preference for interactive reasoning. Some participants preferred to monitor the intermediate process. In particular, P4 mentioned that \textit{``seeing the reasoning is so important, and I have always been wanting to edit the reasoning chain.''} However, other participants suggested that they did not consider it necessary to see the reasoning steps. P9, a movie producer, commented ``\textit{if I'm having a hard time making a decision, I would want to get a few general recommendations. This process seems too logical,''} suggesting that while insights into the reasoning steps may be helpful in some situations, it may depend on users' decision styles and the contexts. 

The requirement to provide feedback to the model may also lead users to doubt the model's performance, or as P9 told us: ``\textit{if the model needs to
confirm with me so many times, I just felt that the model is not that good.}'' The diverging opinions on interaction frequency is in line with our result that there is no significant difference between the ratings on Future Use ($M_{Baseline}=4.00, M_{\interface}=5.25, p = 0.089$). 

\textbf{Trade-off between granular details and high-level summary.}
We developed \interface to uplift the transparency of the intermediate reasoning steps and encourage users to intervene and control reasoning when possible. While many participants appreciate being more attentive to and making sense of the reasoning chains, a few participants commented on the tradeoffs between attending to the granular details and high-level summaries. P11 commented that when a user asks \interface a question, they may not want to spend too much time focusing on the reasoning chain: \textit{``While I read a lot of the reasoning, it might not be practical to do so in every case.''} However, most participants acknowledged that they wanted to double-check the model's reasoning if they were to ask AI for advice for themselves in the real world. 

Zero participants used  ``Collapse the tree'' function while the tree generation unfolded. Only three participants used it to collapse a subtree after the complete tree generation, primarily to revisit the overall reasoning process. When asked during post-study interviews why they did not use this feature during tree generation, participants explained they preferred maintaining full transparency of all reasoning steps on the fly. For instance, P5 mentioned that \textit{``currently I know that I can collapse the tree after the reasoning process completes, but I kind of want to see the response quickly on the fly. Maybe you can try a more top-down approach where users can expand the tree.''} This suggests a potential complementary design from the current depth-first tree traversal to a breadth-first tree traversal, which we further discuss in Section \ref{sec:discussion}.

%% file: 05-case-studies.tex
\section{Case Studies}

After our user study (Section \ref{sec:eval}), two participants requested to use \interface for their real-world use cases. We followed up with the participants to explore other potential use cases and how users interact with reasoning in other decision-making scenarios. We demonstrate how they explore \interface and design insights from these sessions. To distinguish from participants in Section \ref{sec:eval}, we refer to participants as C1 and C2. We display text that participants typed into and nodes on \interface as \query{text against a light gray background}.

\subsection{Information Seeking}

C1, a computational neuroscientist, had recently used an LLM chatbot for a question \query{How does hippocampus consolidate memory back to the neocortex?} for their own research studies. C1 typed and asked this question on \interface, which generated, as C1 commented, a graphical \textit{mindmap} of reasoning.

C1 followed the tree generation. In the first subtree, \interface showed details about a related term \query{systems consolidation}: \query{Maybe during sleep, especially slow-wave sleep, the hippocampus replays memories, which helps transfer them to the neocortex.} One child node followed this topic: \query{There's something about sharp-wave ripples in the hippocampus during this replay.} The user halted the generation and asked their follow-up question (via \ilabel{c} from \autoref{fig:demo}): \query{Does sharp-wave ripple carry memory information?} C1 was satisfied with the response to this small point and stored this in this subtree. Later, C1 also added another follow-up question: \query{What are the circuits that connect hippocampus and entorhinal cortex} when the \interface mentioned \query{neocortex}. C1 commented that the ability to ``\textit{walk through the mindmap with the reasoning is actually very useful}'', and the ability to comment and follow up with the model reasoning made them ``\textit{very engaged in the answer}'' for this use case. 

However, C1 wished to introduce further \textit{control} not only to the individual nodes, but the entire tree generation process. While they acknowledged that the full reasoning chain was \textit{``definitely more comprehensive,''} C1 commented that they ``\textit{kind of got the answer}'' in the middle of the tree generation, and ``\textit{at this point, I really wanted to skip and see the [subsequent final] response, almost like a summary of what I just answered in the process.}'' This indicates future design opportunities to skip subtrees where users are familiar with the subtopics and reveal the final model output promptly.

Another feedback was that the \texttt{Feedback} node interaction is not exactly what they would like to stop. For instance, \interface stopped at a \texttt{Feedback} node for C1's clarification: \query{But I'm a bit fuzzy on the mechanisms of reactivation.} C1 stated ``\textit{I can see why it was stopped, but I think I don't really feel like giving input here}'' because this was not a point of confusion for C1. When asked about the difference between reviewing the reasoning chain and a traditional chatbot interface (e.g., ChatGPT), C1 recognized that reviewing the reasoning ``\textit{is just more engaging for this task which honestly requires a lot of attention for him}''. The right or wrong answer seems less critical, compared to going through different topics in a \emph{mindmap}.

\subsection{Financial Planning}

C2, a financial analyst, had recently been tasked with evaluating the potential acquisition of another company (referred to as company A). They were in the early stage of the evaluation and wanted to navigate through different aspects of this process. They typed their query \query{I worked at [Anonymized Company] in their corporate development team. We are looking to acquire a company called Company X [a brief description of this company]. Help me make the case for why we should or should not acquire this company.} During the interaction, we observed that C2 consistently collapsed each reasoning tree as it was generated, creating a more manageable visual hierarchy. At the \texttt{Feedback} nodes, C2 provided domain-specific expertise in response to clarify questions about Company A's market share and competitive technological positioning. This contextual information was then incorporated into subsequent branches.

Upon completing the session, C2 reflected: \emph{``This is essentially a plan for me to consider. What this really represents is the decision-making process that happens in corporate boardrooms. What I appreciate about this approach, compared to ChatGPT or DeepSeek, is that it effectively models C-suite decision-making processes.''} C2 elaborated that in large organizations, acquisition discussions involve consideration of multiple factors, and this interface allowed them to anticipate these considerations and get ahead of these ideas.

When asked to compare this experience with chatbot interfaces, C2 emphasized the value of depth over speed: \textit{``The text alone in standard responses doesn't provide much nuance or reveal the underlying stream of consciousness. For a financial analyst, it's not about how quickly we can complete this process. I'm focused on how diligently we can review the logic behind the decision.''} C2 further noted that a conventional output listing pros and cons without revealing the reasoning process would prompt immediate questions about the rationale, stating: \emph{``If you just show the pros and cons in the final answer, I would immediately ask why is it so?''}

%% file: 06-discussion.tex
\section{Discussion and Design Opportunities}
\label{sec:discussion}

Our work introduces user interaction with reasoning chains for LLMs. We discuss the implications of our findings as they relate to making reasoning processes transparent and paving a future design space for interaction during test-time scaling.

\textbf{Improving user agency may improve perceived output quality in reasoning models}.
In our user study, all participants valued the transparency of reviewing the reasoning process before seeing the final model output. Reviewing the reasoning helped them understand the tradeoffs of these oftentimes subjective decision-making tasks. Having the direct entry point of interaction on \interface made participants feel their voices are heard. Moreover, participants observed that having more direct control over the reasoning process made the final output personalized. The empirical evidence of an increased sense of control over the reasoning process suggests users benefit from engaging with the reasoning process. While work in the NLP community has emphasized the goal of test-time scaling to improve the model output~\cite{snell2024scaling}, the observations in our study challenge recommendations~\cite{baker2025monitoring, detectingOpenAI} that suggest concealing intermediate reasoning processes from users or limiting user control over the reasoning process. 

\textbf{Re-purposing model output after interactive reasoning}. 
 We found mixed results on participants' caution over the final model output. While prior work in XAI suggests that cognitive forcing functions may reduce overreliance on the model output, our findings add nuances. For high-stakes decision-making tasks, users may consult LLMs for tradeoffs but may defer to themselves to make such decisions. In fact, as P2 put it in Section \ref{sec:repurposing}, users may just gain ``\textit{a better understanding of the situation already}'' by going through the reasoning process. While test-time scaling is now primarily viewed as a means to improve model responses~\cite{baker2025monitoring, snell2024scaling}, in the context of decision-making tasks, the final response could be re-purposed as a personalized summary, rather than the main objective that models typically produce. 
 This finding suggests a potential shift in how we conceptualize the ``output'' of AI systems---from the final objective to supportive reasoning artifacts that enhance users' decision-making. 

 \textbf{Tensions in designing for interactive reasoning}.
 Our study also revealed potential difficulties in how users engage with LLM reasoning. Participants found that the long presentation of reasoning in the baseline created significant cognitive barriers that prevented meaningful engagement, despite participants' expressed interest in understanding the model's thought process. \interface alleviated some of these problems by transforming the ``wall of text'' into an interactive tree representation. We note that participants were more aware of the assumptions from the reasoning by interacting with \interface; after all, making an informed decision requires navigating nuanced trade-offs. However, the tree representation---or even showing any reasoning at all---might be unnecessary for simpler or low-stakes queries. Indeed, some participants wished to bypass the reasoning process altogether after providing some feedback (e.g., C1). This surfaces a design opportunity to calibrate the display of and interaction with reasoning based on task complexity. This might involve adaptive interfaces that present abbreviated reasoning for routine decisions while reserving comprehensive reasoning chains for ones that demand more complex tradeoffs.

 \textbf{Tradeoffs of the depth-first and breadth-first tree structure}.
 \interface was motivated by projects in the UIST community that use a node-link diagram for information sense-making~\cite {jiang2023graphologue, suh2023sensecape}. Our findings (Section \ref{sec:res:control}) of improved sense-making corroborate prior work in the context of reviewing long reasoning during test-time scaling. Meanwhile, we note a potential design opportunity to edit AI reasoning on the fly. 
 Many participants suggested new features to view reasoning at multiple levels of abstraction---first seeing broad conceptual frameworks before selectively exploring specific details of interest (i.e., a \textit{breadth-first} tree traversal). Contrasting this was the \interface design which followed a \textit{depth-first} tree traversal that followed the original trajectory of the reasoning chains. However, as some participants suggested, reasoning involves complex networks of interconnected concepts, premises, and inferences, rather than a purely sequential or linear fashion in current interfaces (including in \interface). 
 The current sequential reasoning paradigm from the reasoning model and the exploration based on hierarchical order reflects a \textit{mismatch between current interface paradigms and the nature of reasoning}~\cite{gross2006starring}.
 Future designs could better support interactive reasoning by adopting visualization approaches that explicitly represent \textit{networked} relationships. This could involve interfaces that support \textit{both breadth-first exploration for context and depth-first exploration for details}, with interactive capabilities to expand or collapse reasoning branches according to user interest. Such approaches could bridge the gap between the linear presentation of current interfaces and the more networked structure of human reasoning.

\textbf{Varied Engagement Preferences in LLM Reasoning}.
We found significant variation in how participants wished to engage with reasoning processes. On the one hand, some participants wished to reduce the engagement level, stating that ``\textit{if the model needs to confirm with me so many times, I just felt that the model is not that good.}'' [P9]
Other participants wish to include even more interaction, such as eliciting users' confirmation when a node branches off. This suggests that future design with model reasoning (e.g., ~\cite{liu2025interacting}) should require considerations such as user's existing level of AI reliance and task complexity. For instance, participants noted that the tree visualization was more valuable for analytical questions than for creative or simple factual tasks. Some participants might prioritize efficiency, while others value learning from the reasoning process itself.
These findings suggest design opportunities for interactive reasoning that accommodates varying levels of engagement across different contexts and decision-making styles~\cite{mei2025passingbuckaiindividuals}. Potential designs that support this variance may range from fully automated reasoning with minimal visibility to highly interactive approaches in which users actively shape the reasoning trajectory. 

\section{Limitations and Future Work}

Limitations of this work are the relatively limited sample size (N=16) as well as participants' relatively high familiarity with LLM models, which may limit the generalizability of our findings to novice populations. We attempted to alleviate this limitation by recruiting participants from diverse backgrounds ranging from undergraduate students to working professionals in finance, computer science, and art. While we used two daily dilemma scenarios to increase the stakes of the task and allow fair comparison between the two conditions in the study, decision-making in knowledge-intensive scenarios (e.g., coding~\cite{yen2024coladder} or medical diagnosis~\cite{szolovits1988artificial}) might require information from beyond just model reasoning.

Another limitation is that our current approach does not systematically analyze the model behaviors given users' feedback. In fact, a recent study~\cite{chenreasoning} of reasoning models during test-time scaling without human feedback shows that CoTs may not faithfully represent a model's actual reasoning process to answer math and coding problems. While our paper does not claim that user feedback led to more personalized or accurate reasoning chains or final model outputs, participants observed that their inputs were incorporated into the final outputs, which demonstrated several benefits from the end user perspective. We also acknowledge that the \emph{Link} operator matches the sentences in the model response and the content within each reasoning node based on semantic similarity; however, this linkage does not indicate a causal relationship internal to the model. Future work may consider investigating more robust approaches to derive this causal relationship for further explainability.

We acknowledge that \interface may incur misuse for ethical consideration. We recognize that LLMs, such as \texttt{gpt-4o} and \texttt{DeepSeek-R1} can hallucinate and generate false information that may affect daily decision makings. Users might also mistakenly steer model output by providing existing biased or malicious feedback, leading to harmful result but became more confident in the end~\cite{sharma2024generative}. To remedy this, we cautioned users for such risks in our system and study. We also chose daily dilemma tasks in the user study rather than topics that can be highly controversial, such as political disagreement~\cite{fisher2024biased}.

\section{Conclusion}
In this paper, we introduced \interface, a system that instantiates \emph{interactive reasoning}, an approach that visualizes the LLM reasoning steps via test-time scaling and allows users to make sense of, and control the reasoning before reaching the model's final output.  We evaluated \interface through a user study with 16 participants from diverse occupational backgrounds, as well as two case studies to explore how users leverage interactive reasoning in decision-making tasks. Results showed that \interface increased the sense of control, sense-making, and assumption in the model output compared to a baseline system. We also observed participants' repurposing of model responses after engaging with the reasoning steps. We discuss practical implications for future adaptive designs that support interaction with a model's reasoning steps. Overall, our contributions set the stage for new interactive paradigms for test-time scaling not just for model response quality but also to improve user control through human agency.

%% file: appendix.tex
\section{An example of DeepSeek-R1 response with the reasoning steps}

\begin{figure}[h!]
    \centering
    \includegraphics[width=\linewidth]{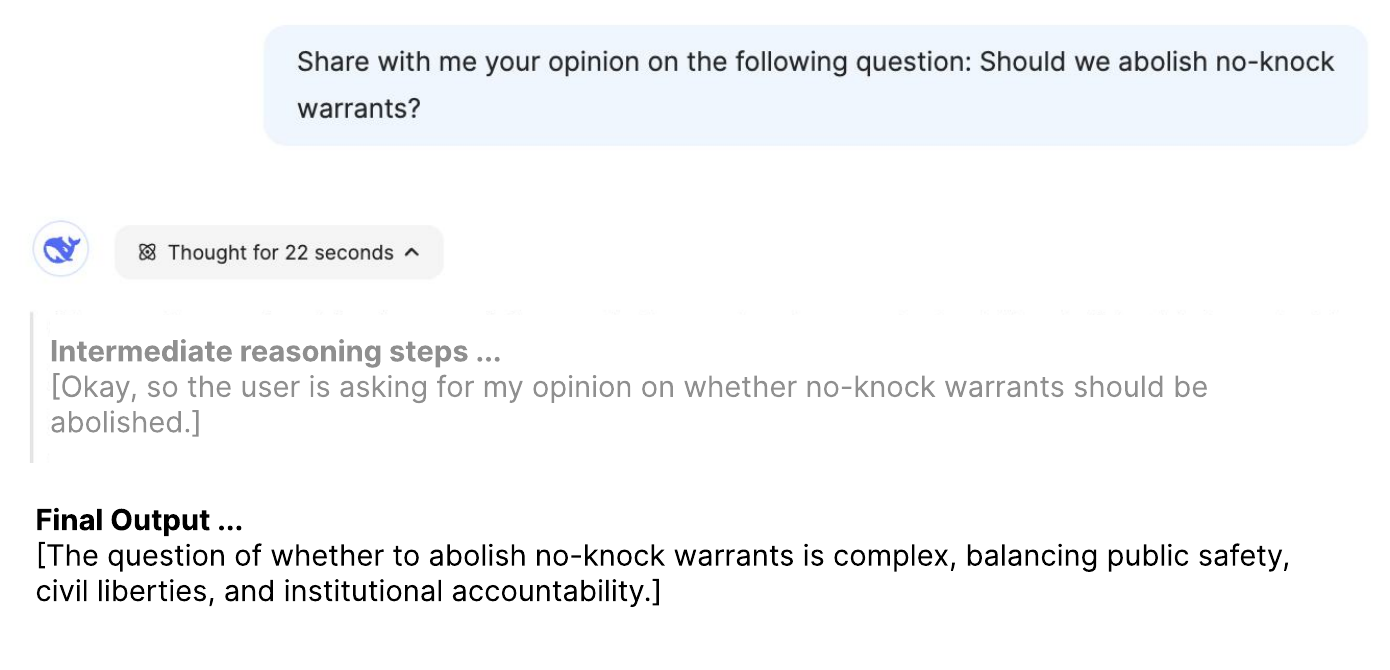}
    \caption{An example of the current DeepSeek-R1 platform that shows the reasoning steps. The model thought for 22 seconds with over 600 words in the reasoning.}
    \label{fig:example}
    \Description{An example of the current DeepSeek-R1 platform that shows the reasoning steps. The model thought for 22 seconds with over 600 words in the reasoning.  }
\end{figure}

\section{The System Prompts for Operators in the Tree Generation Pipeline}

\subsection{The Structure Operator}
\label{appendix:structure}

This prompt structures the original chain-of-thoughts (CoTs) reasoning into sub-topics. Before applying this operator, we grouped the thoughts first (\ilabel{1} Figure 3) into manageable topics, in line with prior work to leverage LLMs to aggregate concepts~\cite{lam2024concept}. In the project, we observed that CoTs during test-time scaling often lead to many paragraphs, which made the tree very sparse/shallow without this step. Below are the prompts for both the \emph{Structure} operator and the group thought pre-processing.

\subsubsection{Structure}

\begin{code}
You are a helpful assistant that *only* tags the chain of thought for a given text. 
    
Rule 1: Use <topic>...</topic> to indicate a major new area of thought.

Rule 2: Use <branch>...</branch> to indicate a subtopic extending from a previous point.

Rule 3: Nesting Structure:
    1. All content must be contained within tags, with no unmarked text.
    2. <topic> tags should only appear at the top level.
    3. <branch> tags can nest inside <topic> tags or other <branch> tags.
    4. Each <branch> must begin with at least one complete sentence before any nested branches.
Please follow the rules strictly.

    Example:

    \begin{topic}
    Okay, the user is asking where they should travel during spring break. Hmm, I need to consider different types of destinations to cover various interests.
    \end{topic}
    
    \begin{topic}
    First, I should consider different destinations. I should list options that cater to different preferences.
    \end{topic}  
    
    \begin{branch}
        \begin{branch}
            For beach destinations, places like Cancun, Miami, or the Caribbean come to mind.
            \begin{branch}
            These spots are popular for spring break because they offer warm weather, beaches, and vibrant nightlife.
            \end{branch}
        \end{branch}
        \begin{branch}
            But maybe I should also include some less crowded beaches for those who prefer a quieter time.
            \begin{branch}
                Costa Rica is known for eco-tourism, rainforests, and activities like zip-lining.
            \end{branch}
            \begin{branch}
                Hawaii is another option with hiking and volcanoes.
            \end{branch}
        \end{branch}
        \begin{branch}
            Cities with cultural attractions could be another category: places like Paris, Kyoto, or Barcelona. Oh, Kyoto in spring would have cherry blossoms, that's a big plus.
        \end{branch}
    \end{branch}
    \begin{branch}
        Wait, what is the user's budget? I should include some budget-friendly options too.
    \end{branch} [More of this example]
\end{code}

\subsubsection{Group Thoughts}
\begin{code}
    I have this monologue, representing my reasoning for the query: \$\{query\}. Structure this text into high-level themes.

    Rule 1: Keep the exact text from the input text!! Use the same words. Each theme should surround a high-level idea.
    
    Rule 2: New line to separate each theme. The output should be a direct division of the input text into themes under 8 paragraphs.
    
    Rule 3: The output should not include anything beyond the origin input text, or any summary. I'd like to the exact same text from the input.

  Input:
  \$\{reasoning.replaceAll("\textbackslash n", "")\}

  Output: 
\end{code}

\subsection{The Clarify Operator}
\label{appendix:clarify}

\begin{code}
You are a helpful assistant that *only* tag the chain of thought (which was generated by a model) for a given text. The goal is for users to help clarify the uncertain or incorrect assumptions in the input reasoning chain. You use <user></user> to tag such text.

Rule 1: Identify places where user input would be valuable (uncertainty, preferences, personal experiences)
    
Rule 2: Sentences like "I don't know X" that the reasoning chain is unsure about the situation.

Rule 3: Preserve the original text from the input and only add the <user> tag to the sentences that need clarification.

Rule 4: A good user should tag an uncertain question or a situation so that a user can easily give their feedback or context. 

Rule 5: Do not tag questions that are answered in the reasoning chain later in the text.

Rule 6: The user tag should only appear between <branch> tags; no user tag between <topic> tags allowed.

Bad example: <branch>Hmmm. Let me think. </branch>
    
Good example: 
<branch><user>But then, how to enforce that? It's tricky because everyone has different schedules. What rules would be reasonable to create and enforce in this situation?</user></branch>

\begin{branch}
    \textbf{<user>}Wait, what is the user's budget? I should include some budget-friendly options too.\textbf{</user>}
\end{branch} [More of this example]

\end{code}

\subsection{The Link Operator}
\label{appendix:link}

\begin{code}
    Given the following premises (reasoning nodes) and hypotheses (response paragraphs), determine the entailment relationship between them.

    PREMISES:
    [
    
    \{{"id": \$node\_1\_id, "content": \$node\_1\_content}\}, 
    
    \{{"id": \$node\_2\_id, "content": \$node\_2\_content}\}, ...]
    
    HYPOTHESES:
    [
    
    \{{"id": \$response\_1\_id, "content": \$response\_1\_content}\}, 
    
    \{{"id": \$response\_2\_id, "content": \$response\_2\_content}\}, ...]
    
    For each hypothesis (response paragraph), identify the premise that most strongly entails or supports it. Consider the semantic and logical relationship between each premise-hypothesis pair.
    
    Return your analysis as a valid JSON array with objects containing:
    
    \{
    
      \hspace*{2em}"hypothesis\_id": [response ID number],
      
      \hspace*{2em}"entailing\_premise": \{
      
        \hspace*{2em}"premise\_id": [most relevant node ID],
        
        \hspace*{2em}"entailment\_strength": [confidence score between 0 and 1]
        
      \hspace*{2em}\}
      
    \}
    
\end{code}

\subsection{Other Prompts}

\interface also has a summarization feature to collapse the (sub)-trees. This operator aggregates the subtree nodes into a low-level context, then performs a summarization with \texttt{GPT-4o}. Note that when users expand the summarized subtree, the child nodes remain the same and are not regenerated. 

\begin{code}
    Given the context, please summarize these thoughts into a paragraph of summary under 60 words.

  Content: \$\{subtree\_context\}
  One sentence summary:
\end{code}